\documentstyle[aps,epsf]{revtex}  

\def\Psibar{\overline{\Psi}}

\def\meff{m_{\rm eff}}

\def\chidof{\chi^2 / {\rm d.o.f.}}

\def\spose#1{\hbox to 0pt{#1\hss}}
\def\ltapprox{\mathrel{\spose{\lower 3pt\hbox{$\mathchar"218$}}
 \raise 2.0pt\hbox{$\mathchar"13C$}}}
\def\gtapprox{\mathrel{\spose{\lower 3pt\hbox{$\mathchar"218$}}
 \raise 2.0pt\hbox{$\mathchar"13E$}}}
\def\inapprox{\mathrel{\spose{\lower 3pt\hbox{$\mathchar"218$}}
 \raise 2.0pt\hbox{$\mathchar"232$}}}

\begin{document}        

\baselineskip 14pt
\title{Dynamical lattice QCD thermodynamics and the 
$U(1)_A$ symmetry with domain wall fermions}
\author{Pavlos Vranas for the Columbia lattice group
\thanks{P.~Chen,
N.~Christ,
G.~Fleming,
A.~Kaehler,
C.~Malureanu,
R.~Mawhinney,
G.~Siegert,
C.~Sui,
P.~Vranas,
L. Wu,
Y.~Zhestkov
}}
\address{University of Illinois at Urbana-Champaign, Urbana, IL 61801}
%
\maketitle              

\begin{abstract}        
Results from numerical simulations of full, two flavor QCD
thermodynamics at $N_t=4$ with domain wall fermions are presented. For
the first time a numerical simulation of the full QCD phase transition
displays a low temperature phase with spontaneous chiral symmetry
breaking but intact flavor symmetry and a high temperature phase with
the full $SU(2) \times SU(2)$ chiral flavor symmetry. It is found that
close to the transition in the high temperature phase the $U(1)_A$ axial
symmetry is broken only by a small amount. This result is of particular 
interest because of the connection between $U(1)_A$ symmetry breaking 
and the expected order of the transition.
\
\end{abstract}   	

\section{Introduction}               
\label{sec_intro}

The upcoming RHIC experiments will reproduce the extreme conditions
that existed in the early universe and it is hoped that the predicted
QCD phase transition from hadronic matter to a quark-gluon plasma will
be observed. The non-perturbative nature of these phenomena makes
precise theoretical predictions difficult.  Lattice gauge theory
provides a first-principles approach to QCD thermodynamics and, in
principle, a way to perform precise non-perturbative calculations of
thermodynamic quantities at equilibrium. Quantities such as the value
of the critical temperature, the width of the critical region and the
order of the transition can then be used by phenomenological models to
make contact with experiment. While great progress has been made since
the inception of the lattice regulator 25 years ago \cite{Wilson}
there remain very significant uncertainties that prevent these
calculations from being viewed as unambiguous predictions of QCD.
Since these calculations are computer simulations one could
think that progress in this field is tied with progress in computer
speeds. While this is partially true, it is not the real culprit
especially given the fast growth in computer speeds during this
period. The true reason lies at the most simple and fundamental level
of the lattice theory.

When the first derivative of the fermion kinetic energy term is made
into a lattice difference operator the number of light species
multiplies by a factor of $2^d$ where $d$ is the dimension of
space--time. This is the well known doubling problem and it turns out
that at non--zero lattice spacing in order to remove the extra species
and still have a local Lagrangian one has to compromise the global
flavor chiral symmetries of the theory \cite{Nielsen_Ninomiya}. There
are two popular methods to put fermions on the lattice: Wilson and
staggered fermions.  Both of these methods at finite lattice spacing
$a$ break the chiral symmetry.
%
%
The chiral symmetry is recovered together with the Lorentz symmetry as
the continuum limit $a\rightarrow 0$ is taken.  And here lies the
problem. In a numerical simulation a decrease of the lattice spacing
by a factor of $2$ requires an increase in computation by a factor of
$2^{8-10}$ depending on the parameters.  In order to control the
amount of chiral symmetry breaking induced by the regulator
prohibitively large computing resources are required.  Clearly faster
computers although one day may be able to overcome these large factors
can not be the answer.

A way out of this forbidding state of affairs came from
D. Kaplan\cite{Kaplan} through an unexpected avenue. He formulated a
new fermion lattice regulator with the name domain wall fermions
(DWF).  The first numerical simulations using DWF in vector gauge
theories were done in \cite{PMV} and were followed by
\cite{Blum_Soni}, \cite{lat98_PMV}, \cite{lat98_RDM},
\cite{lat98_Fleming}, \cite{lat98_Kaehler}, \cite{ICHEP_NHC}. In this
work the latest results from simulations of dynamical QCD
thermodynamics are presented.  Preliminary results of this work have
already appeared in \cite{lat98_PMV}, \cite{ICHEP_NHC}.  For promising
alternatives to domain wall fermion simulations see
\cite{Neuberger_fermions}.

\section{Domain wall fermions.}
\label{sec_dwf}

Domain wall fermions introduce an extra direction of space--time with
size $L_s$. The fermion fields are five-dimensional while the gauge
fields remain four-dimensional and couple the same way to all extra
fermion degrees of freedom.  The boundary conditions along the fifth
direction are free (domain wall) and although the extra fermion
degrees of freedom are heavy a light Dirac fermion surface mode
develops on the boundaries with its positive chiral components
exponentially bound on one wall and its negative components on the
other. In the $L_s \rightarrow \infty$ limit the mode is a massless
Dirac fermion. At finite $L_s$ the residual mixing introduces an
exponentially small mass. Since these properties are maintained at any
lattice spacing DWF provide a way of separating the recovery of the
chiral symmetry from the recovery of the Lorentz symmetry.  Now at a
fixed lattice spacing the chiral symmetry can be recovered by
increasing $L_s$.  Unlike Wilson or staggered fermions the computing
cost in recovery of the chiral symmetry is only linear in $L_s$! These
remarkable properties may bring within reach physically realistic
studies of lattice QCD thermodynamics.

The subject of DWF has a large volume of work behind it.  The reader
is referred to \cite{PMV} and references therein for an
introduction. Here, the version of DWF as presented in
\cite{Furman_Shamir} with the modifications in \cite{PMV} is used.
The theory has five parameters: the four-dimensional lattice volume
$V$, the inverse gauge coupling squared $\beta$, a bare mass that
explicitly mixes the two chiralities bound on the domain walls $m_f$,
the number of sites of the fifth direction $L_s$, and a
five-dimensional mass that can be thought of as the domain wall
height $m_0$.  The parameters $m_f$, $L_s$ and $m_0$ set the bare
quark mass.  In free theory the bare quark mass $\meff$ is given by
\cite{PMV}:
\begin{equation}
\meff = m_0 (2 - m_0) \left[ m_f + (1-m_0)^{L_s}\right], \ \ \ \ 0 < m_0 < 2 
\label{meff_free}
\end{equation}

There are many important theoretical issues regarding DWF. Perhaps the
most crucial one relates to the localization of the chiral modes in
the interacting theory.  Does the localization remain exponential as
in free theory or does it become power law? How does the decay rate
depend on the lattice spacing? These questions were addressed in
detail in \cite{PMV} in the context of the two flavor Schwinger model.
There it was found that the decay is exponential and that the decay
rate becomes faster as the continuum limit is approached. The last
result can be understood by a simple examination of
eq. \ref{meff_free}. In the interacting theory $m_0$ will be
renormalized and therefore optimum localization will be shifted away
from $m_0=1$.  However, even if $m_0$ is adjusted to absorb this
renormalization, optimal decay will not be possible since the optimal
value for $m_0$ will fluctuate from one background gauge field to the
next.  At strong coupling were the gauge field fluctuations are large
the corresponding fluctuations in the optimal value of $m_0$ will also
be large and the decay rate will become slow. This scenario is
reminiscent of the way the bare mass of Wilson type fermions is used
to cancel the renormalization of the interacting theory. However,
because the massless point changes from configuration to configuration
masslesssness can only be achieved up to order of the lattice
spacing. Here the same is true but the fluctuations in mass are raised
to the $L_s$ power and therefore can be de-amplified by any amount for
large enough $L_s$.

There is another related source of de-localizing effects. It has
been shown that the transfer matrix along the extra direction
for certain gauge field configurations can develop
unit eigenvalues \cite{NN1}. For such configurations the surface
states completely de-localize. Although, the set of such
configurations is of zero measure, configurations in their
vicinity will produce poor localization. These configurations
are special to the lattice since they allow the passage
from one topological sector to another. Again, as the continuum
limit is approached these configurations will have a negligible
Boltzmann weight and their de-localizing effects will diminish.

Another issue relates to the positivity of the transfer
matrix along the fifth direction. Details regarding this can be found
in \cite{dyn_QCD}. Here, it suffices to mention that if $L_s$ is kept
even and if all Green's functions are extracted from sites along the
fifth direction that have a distance from the boundaries of an even
number of sites the relevant transfer matrix is the square of the
single step one and is therefore positive.

Finally the number of light flavors of the free theory depends on
$m_0$. For example if $0<m_0<2$ the theory has one light flavor
but if $2<m_0<4$ the theory has four light flavors. These $m_0$
ranges are modified in the interacting theory and one should
make sure that the needed number of flavors is produced
\cite{lat98_RDM}, \cite{lat98_PMV}.

\section{Two flavor Schwinger model.}
\label{sec_schwinger}

\begin{figure}[ht]	
\vskip -1.5 cm
\centerline{\epsfxsize 2.6  truein \epsfbox{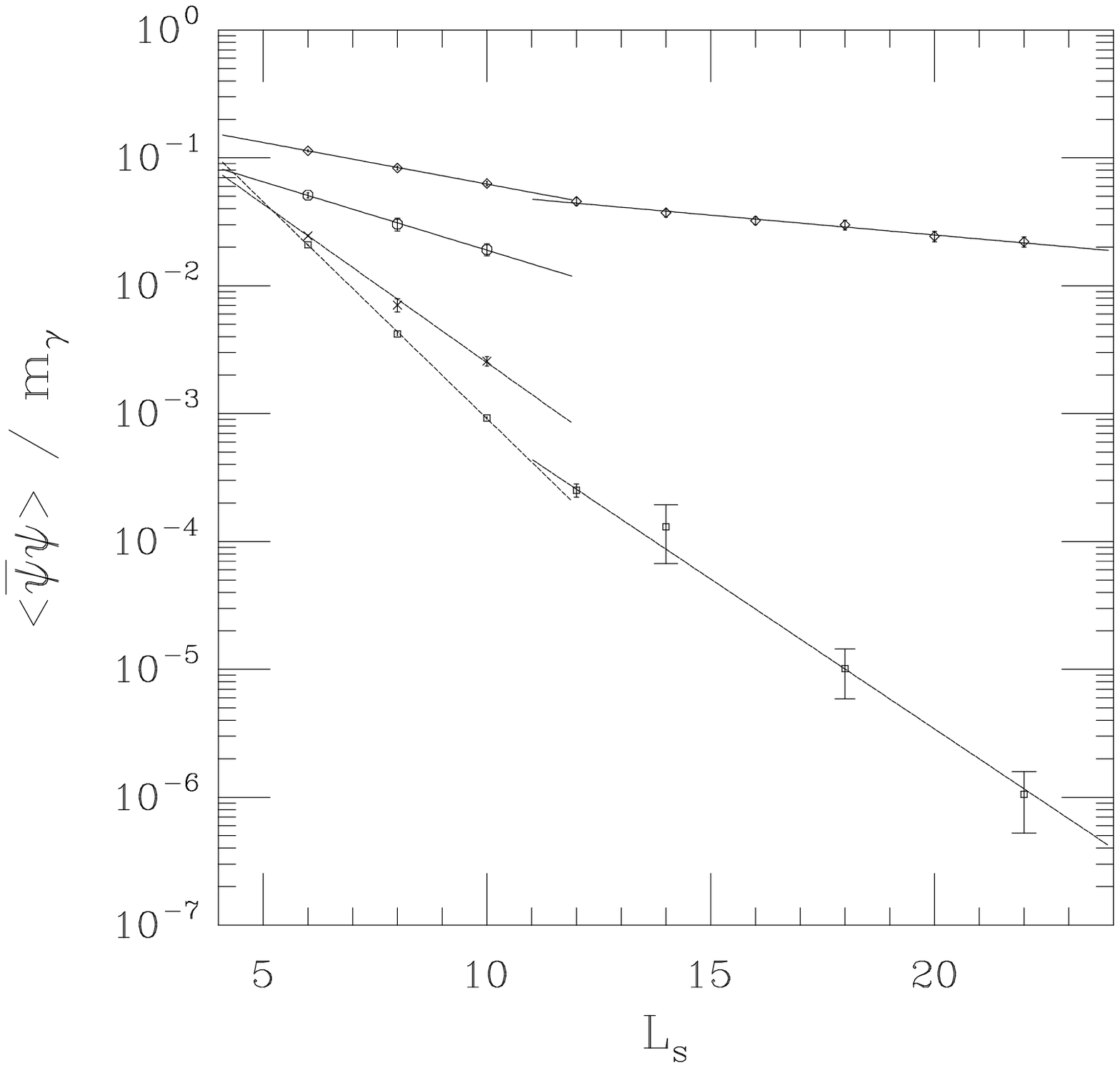}}   
\vskip -.2 cm
\caption[]{
\label{fig_1}
\small The chiral condensate normalized by the photon mass
for $m_f=0$, $m_0=0.9$.
From top to bottom the lines correspond to lattice spacing $a$ of
$1/6, 1/8, 1/10$ and $1/12$. The solid lines are fits to $c_0 e^{c_1 L_s}$.}
\end{figure}

The two flavor Schwinger model was simulated numerically in
\cite{PMV}. If the chiral symmetries
of the massless model are intact, the chiral condensate
will have a zero vacuum expectation value, $<\Psibar\Psi> = 0$
at all couplings.
Therefore, the deviation of $<\Psibar\Psi>$ from zero 
can serve as a measure of 
the amount of chiral symmetry breaking induced by the regulator.
In figure \ref{fig_1} the results from these simulations are presented.

There are three observations that can be made from this figure: \hfill\break
1) For each lattice spacing there are two different decay rates.
In \cite{PMV} the fast rate was related to the physics of the zero
topological sector while the slow rate was related to the physics of
topology changing configurations. \hfill\break
2) Fits to an exponential decay form $c_0 e^{-c_1 L_s}$ have
$\chi^2$ per degree of freedom around one indicating that
the localization of the light states in the interacting theory 
is consistent with exponential decay. Also, power law behavior 
can be excluded with some confidence for the $a=1/12$ data since
a power law fit has a $\chi^2$ per degree of freedom of $32$. \hfill\break
3) The decay rates become faster as the lattice spacing is decreased.
This is a key feature that makes DWF a viable lattice fermion regulator.

\section{Quenched QCD.}
\label{sec_quenched}

Before DWF can be used in QCD an analysis similar to
the one for the Scwhwinger model must be done in order to map the parameter space.
Below the results from such an analysis using quenched QCD are presented.
Preliminary results can be found in \cite{lat98_RDM}, \cite{lat98_Fleming}.

\begin{figure}[ht]	
\vskip -0.5 cm
\centerline{\epsfxsize 4.4 truein \epsfbox{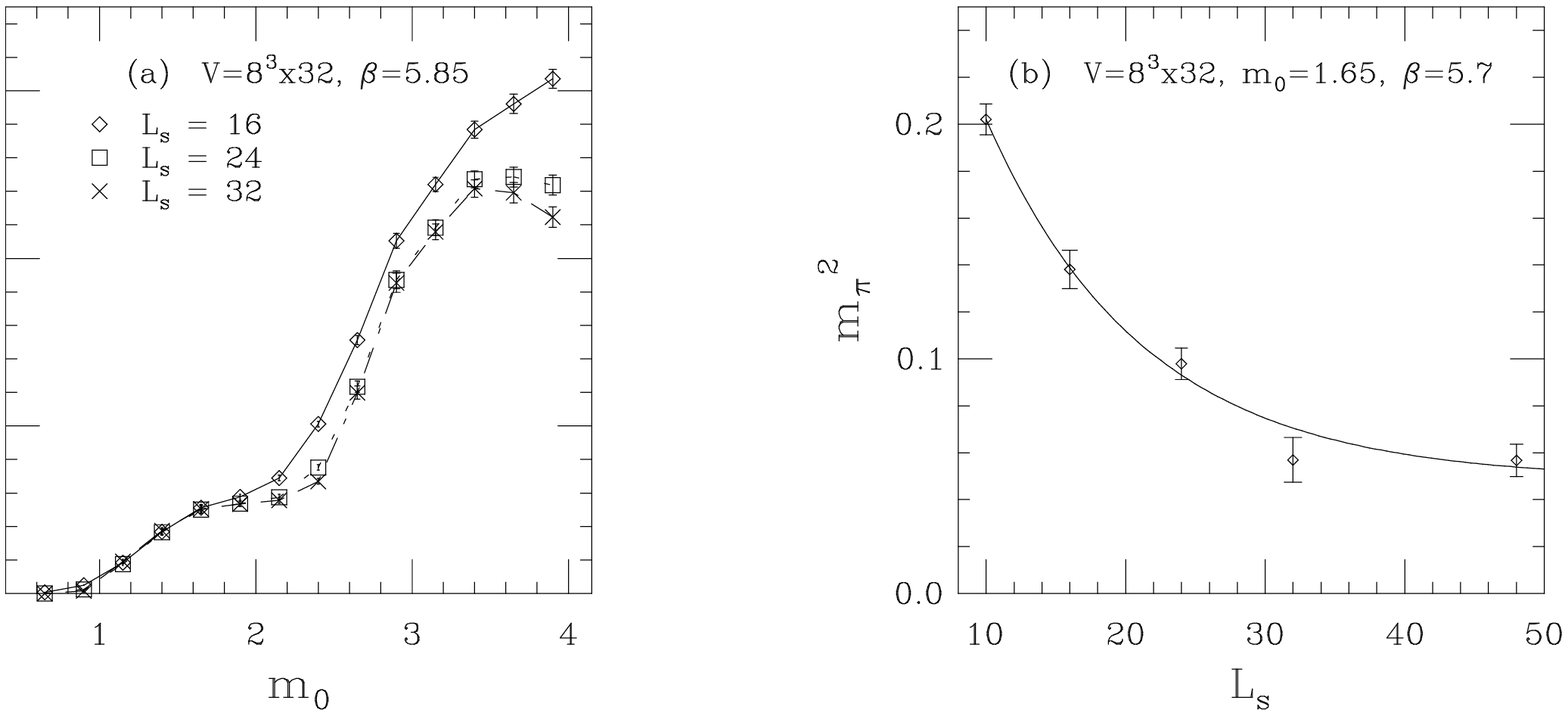}}   
\vskip -2.0 truein 
\caption[]{
\label{fig_2}
\small (a) $<\Psibar \Psi>$ extrapolated to $m_f=0$ vs. $m_0$ for various 
$L_s$. (b) $m_\pi^2$ extrapolated to $m_f=0$ vs. $L_s$.}
\end{figure}

As mentioned in section \ref{sec_dwf} $m_0$ controls the number of
light flavors.  Since this parameter is renormalized, it is important
to understand which ranges correspond to different flavor sectors. In
figure \ref{fig_2}a $<\Psibar\Psi>$ is plotted as a function of
$m_0$. Three regions can be distinguished: $<\Psibar\Psi>$ is zero for
$m_0 < 1$, it has a value around $0.001$ for $1< m_0 < 2.5$ while it
has a value that is four times larger for $2.5 < m_0 < 4.0$ .
Therefore, these three regions correspond to a zero, two and eight
flavor theory (for algorithmic reasons only an even number of flavors
is simulated; had it not been for this restriction the three regions
would correspond to zero, one and four flavors).

In order to investigate the $L_s$ dependence of the theory the pion
mass $m_\pi$ is measured for several values of $m_f$ at fixed $L_s$
and $m_0$ for $\beta=5.7$ on an $8^3 \times 32$ lattice.  Its value is
then extrapolated to zero quark mass. This exercise is repeated for
several values of $L_s$. The $m_f=0$ extrapolated value as a function
of $L_s$ is shown if figure \ref{fig_2}b for $m_0=1.65$. The solid line
is a fit to $c_0 + c_1 e^{- c_2 L_s}$. The fit has $\chidof \approx 1$
indicating that the localization is consistent with exponential
decay. The constant part of the fit is due to finite volume. On a
finite box of size $L$ and periodic boundary conditions the longest
correlation length is $L/2$ and therefore the smallest mass should be
$\approx 2/L$. In this case $m_\pi(m_f=0, L_s=\infty) \approx 0.22
\approx 2/8$.  That this is indeed a finite volume effect was
supported by a second simulation on a $16^3 \times 32$ lattice.

Finally, the rho and nucleon masses were measured for several $m_f$
and $L_s$.  Their $m_f=0$ extrapolated values showed almost no
dependence on $L_s$ for $10 \leq L_s $.  The nucleon to rho mass ratio
was found to be $m_N / m_\rho = 1.508(65)$ to be compared with the
physical value $1.221(2)$. The discrepancy is expected since the
lattice spacing is rather large, $a \approx 0.2 fm$, and the box size
only marginally large $1.6 fm$. Given the modest volume size these
results compare favorably with previous quenched QCD results using
Wilson and staggered fermions at larger volumes.

\section{Dynamical QCD thermodynamics.}

\begin{figure}[ht]	
\vskip -0.5 cm
\centerline{\epsfxsize 4.4 truein \epsfbox{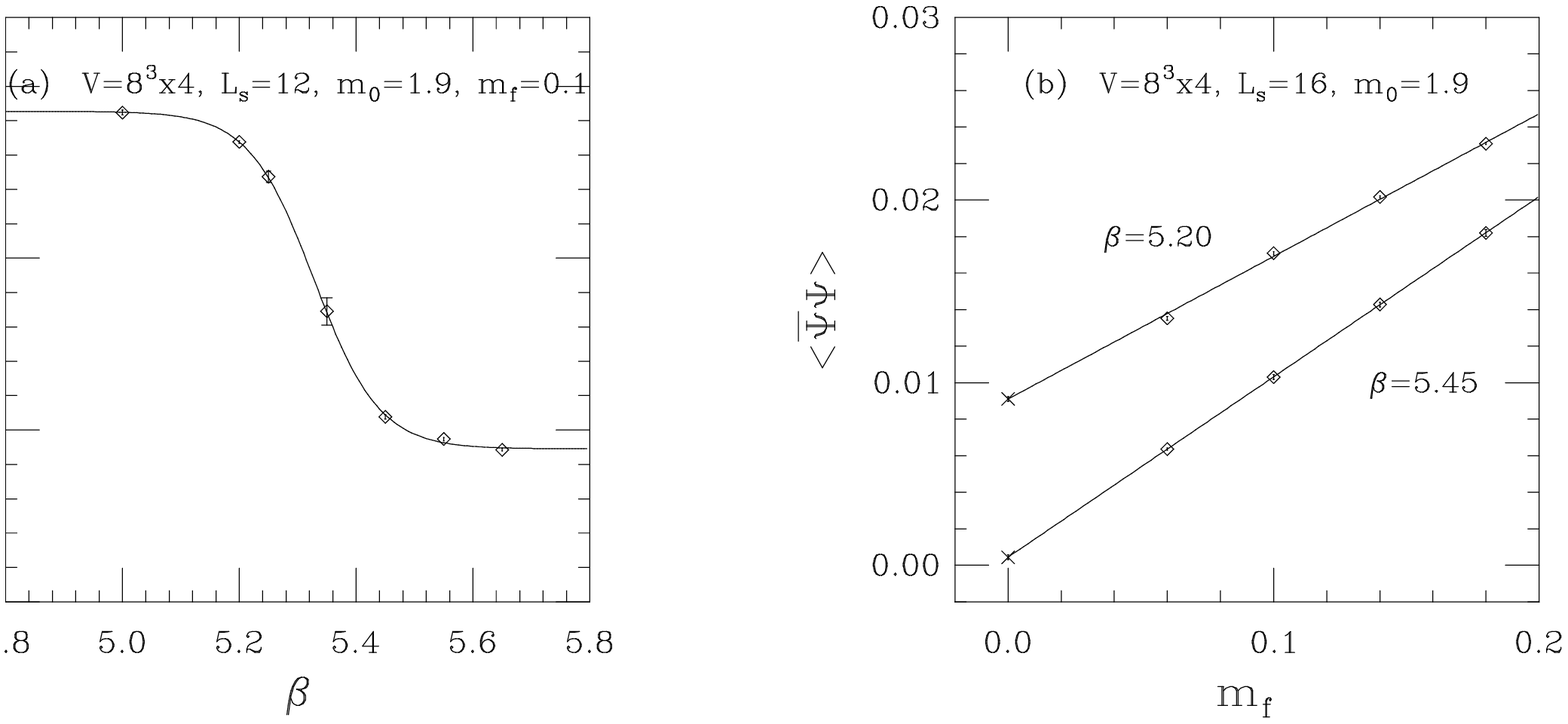}}   
\vskip -2.0 truein 
\caption[]{
\label{fig_3}
\small $<\Psibar \Psi>$ around the QCD finite temperature phase transition.}
\end{figure}

With this ground work in hand dynamical QCD thermodynamics can be
investigated. The first exploratory work is done on small lattices of
size $8^3 \times 4$. Preliminary results have been presented in
\cite{lat98_PMV}. In figure \ref{fig_3}a $<\Psibar \Psi>$ as a function
of $\beta$ is shown for a fixed $L_s$ and $m_f$ at $m_0=1.9$.  A rapid
crossover is observed indicating the presence of a phase transition.
To confirm that indeed there are two phases $<\Psibar \Psi>$ is
plotted in figure \ref{fig_3}b as a function of $m_f$ at $L_s=16$ for 
$\beta=5.20$ below the transition and $\beta=5.45$ above. 
Clearly below the transition
$<\Psibar \Psi>$ extrapolates to a non zero value indicating
spontaneous breaking of the $SU(2) \times SU(2)$ chiral symmetry down
to $SU(2)$ flavor.  Because the Langrangian is explicitly symmetric
under the $SU(2)$ flavor there are three
degenerate Goldstone pions.  
%
%
Above the transition $<\Psibar \Psi>$ extrapolates to a non zero but
very small value (which could be further reduced by
increasing $L_s$) indicating that the full $SU(2) \times SU(2)$ chiral
symmetry is essentially restored. This illustrates the unique
properties of DWF.

\begin{figure}[ht]	
\vskip -0.5 cm
\centerline{\epsfxsize 4.4 truein \epsfbox{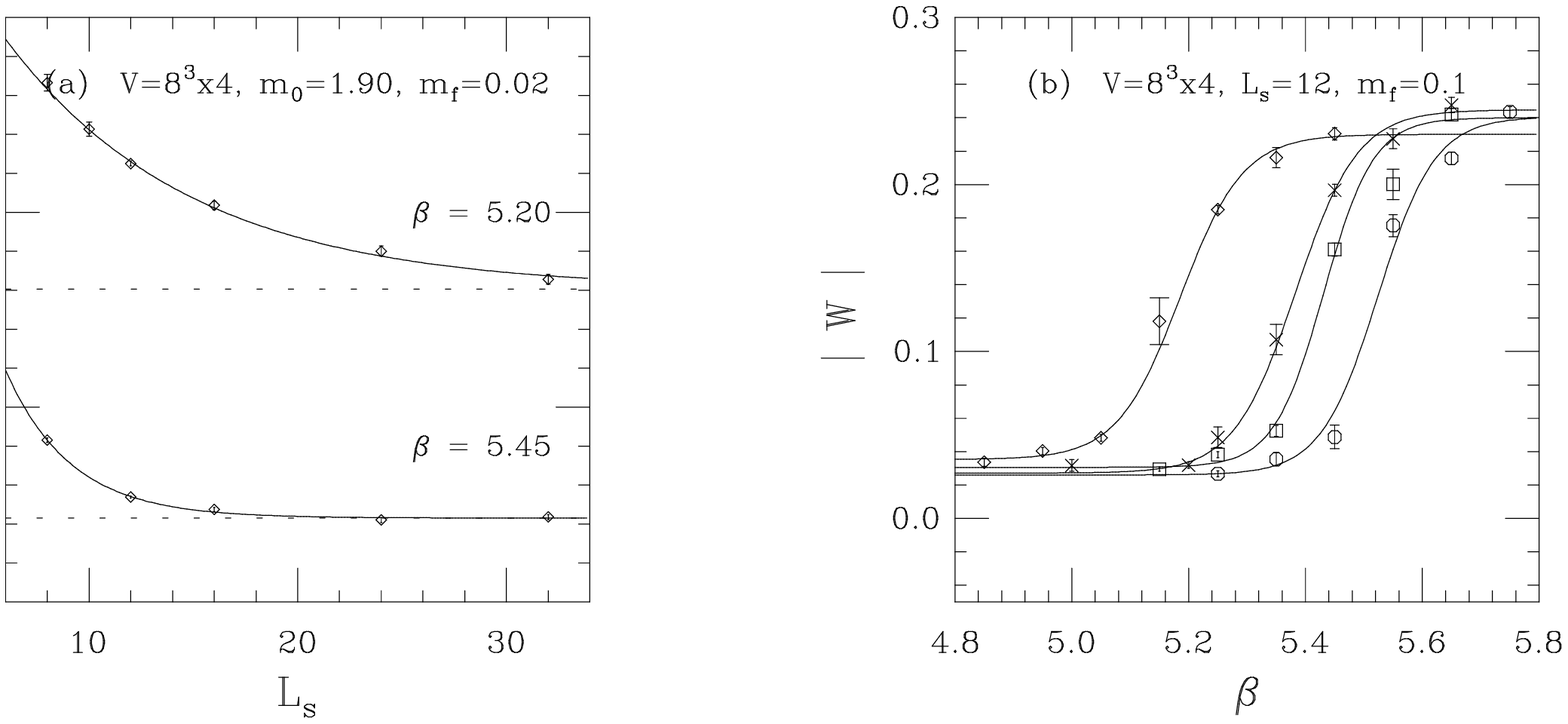}}   
\vskip -2.0 truein 
\caption[]{
\label{fig_4}
\small (a) $<\Psibar \Psi>$ vs. $L_s$ below and above the transition.
(b) The Wilson line $|W|$ vs. $\beta$ for various $m_0$.}
\end{figure}

To further study the properties of the dynamical theory $<\Psibar
\Psi>$ is plotted as a function of $L_s$ for $m_f=0.02$, $m_0=1.9$
both below ($\beta=5.20$) and above ($\beta=5.45$) the transition in
figure \ref{fig_4}a.  The fits are of the form $c_0 + c_1 e^{-c_2
L_s}$. The constant term is the value of the condensate at the
non-zero mass $m_f$.  The $\chidof$ is small indicating that the
localization of the dynamical theory is consistent with exponential.
As seen in the Schwinger model the decay rate becomes faster as the
lattice spacing is made smaller (increasing $\beta$).  At $\beta=5.45$
the $L_s=\infty$ value of $<\Psibar \Psi>$ is already approached to
within a few percent at $L_s=16$ while at $\beta=5.20$ a larger $L_s$
is needed.  Although these results are encouraging it must be stressed
out that at this large lattice spacing other observables may have
different decay characteristics.

Finally it is important to investigate the behavior of the phase
transition as a function of $m_0$. In figure \ref{fig_4}b the Wilson
line is shown as a function of $\beta$ for various $m_0$.  A similar
plot has also been obtained for $<\Psibar \Psi>$ but because the
wavefunction renormalization of the fermion field depends on $m_0$,
it is hard to display in a single figure.  From left to right the figure
corresponds to $m_0 = 2.15, 1.90, 1.80, 1.65$. The
qualitative features of these curves are similar but the critical
value of $\beta$ is a strong function of $m_0$. This is not surprising since
$\beta$ is a bare parameter. One observes that as $m_0$ is reduced the
critical $\beta$ approaches the quenched value $\beta_c = 5.7$
indicating that the regime of zero flavors is approached. The role of
$m_0$ in thermodynamics should be studied with some caution. For
example, from free field studies the range of small momenta accessible to
the light flavor is restricted if $m_0$ is set close to its lower
bound for one flavor physics.  Similar effects can occur if $m_0$ is
set close to its upper bound for one flavor physics.  Since such a
restriction may affect the thermodynamics one should set $m_0$
somewhere in the middle of the allowed range.

\section{The $U(1)_A$ symmetry just above the QCD deconfining transition.}

\begin{figure}[ht]	
\vskip -0.5 cm
\centerline{\epsfxsize 4.2 truein \epsfbox{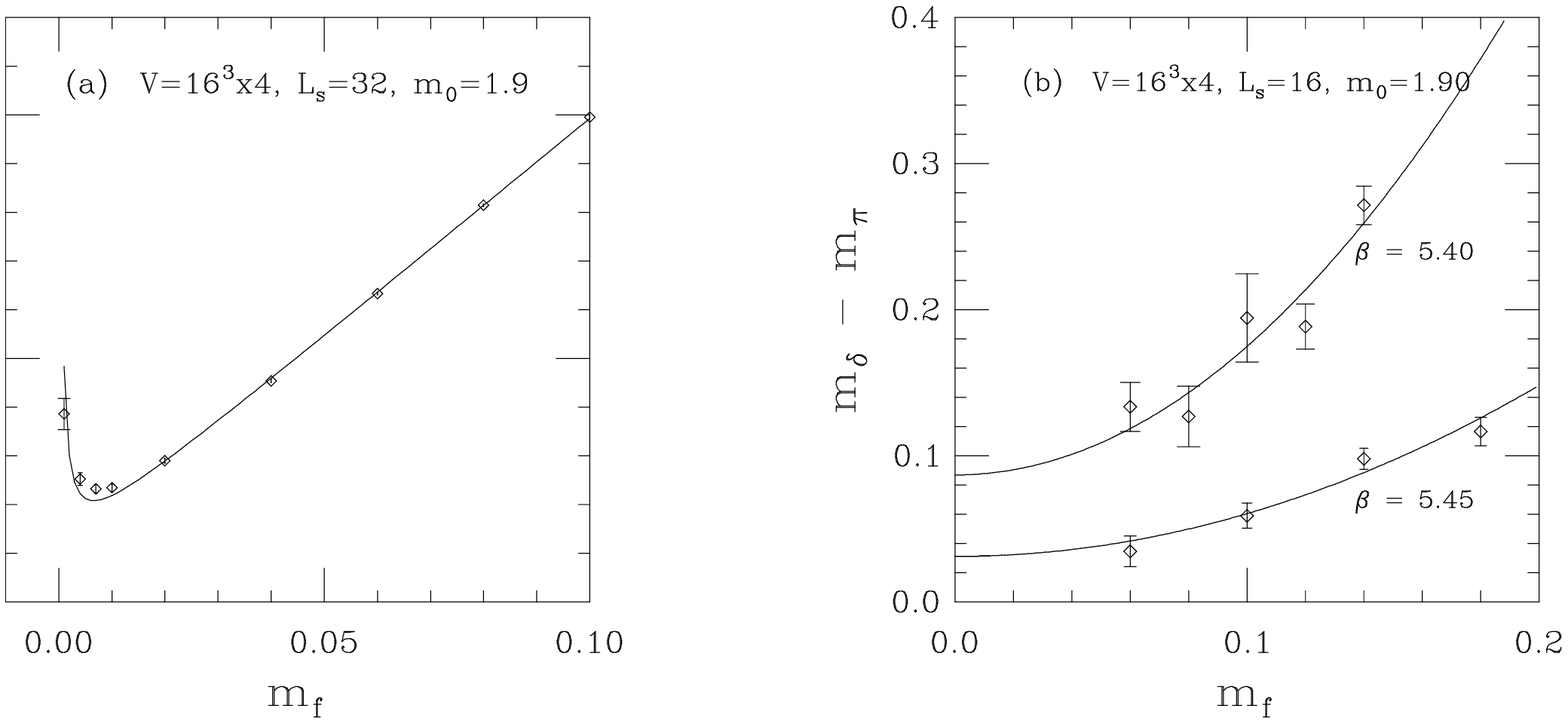}}   
\vskip -2.0 truein 
\caption[]{
\label{fig_5}
\small (a) $<\Psibar \Psi>$ in quenched QCD just above the transition.
(b) $m_\delta - m_\pi$ in dynamical QCD just above the transition.
}
\end{figure}

If the $U(1)_A$ symmetry above but close to the deconfining phase
transition is broken the transition is expected to be second order
while if it is not broken or only softly broken the transition may be
first order \cite{Pizarski}.  This important issue can be investigated
by direct lattice QCD simulations just above the transition. Using
staggered fermions \cite{omega} it was not possible to draw unequivocal
conclusions mainly because staggered fermions may not produce the zero
mode effects necessary for anomalous breaking of $U(1)_A$
\cite{Kogut_Lagae_Sinc}, \cite{Kaehler}.  On the other hand DWF are
known to posses exact and robust zero modes in the $L_s=\infty$ limit
\cite{NN1}. In \cite{Columbia_1} it was shown that for classical
instanton backgrounds these properties are maintained with high
accuracy even at finite $L_s \approx 10$.  In order to confirm the
presence of zero modes in a situation with quantum fluctuations
$<\Psibar \Psi>$ versus $m_f$ from a numerical simulation of quenched
QCD just above the transition is plotted in figure \ref{fig_5}a.  Since
in the quenched approximation the fermionic determinant is set to one,
the zero mode effects are not suppressed and $<\Psibar \Psi>$ diverges
as $1/m_f$.

Having established that DWF exhibit the desired zero mode effects the
$U(1)_A$ symmetry is probed by numerical simulations of full QCD using
DWF on a large lattice above the deconfining transition.  The
difference $m_\delta - m_\pi$ of the screening masses of the delta and
pion particles is used as a measure of anomalous symmetry breaking.
The screening masses are measured from the exponential fall-off of the
relevant two point Green's function along a spatial direction.  Since
the delta and pion Green's function are related by a $U(1)_A$
transformation $m_\delta - m_\pi$ should be zero in the zero mass
limit if the symmetry is not broken and non-zero otherwise.  In figure
\ref{fig_5}b $m_\delta - m_\pi$ versus $m_f$ is shown for
$\beta=5.45$ and $\beta=5.40$ on a $16^3 \times 4$ lattice with
$L_s=16$ and $m_0=1.9$.  The critical $\beta$ is around $5.325$. The
lines are fits to $c_0 + c_2 m_f^2$ and have $\chidof
\approx 1$. The absence of a linear term indicates that chiral
symmetry is effectively restored. The $m_f=0$ extrapolated values
are $0.087(17)$ at $\beta=5.40$ and $0.031(9)$ at $\beta=5.45$.  Although
both are not zero their value is small when compared with
$m_\delta$  and $m_\pi$ which are $\approx 1.3$. 
Universality arguments require that if the QCD phase transition
is to be second order, the anomalous $U(1)_A$ must be broken.
It is an open question as to whether the small size of the $U(1)_A$
symmetry breaking seen here is sufficient to support this theoretical
prediction that the two-flavor QCD phase transition is second order
\cite{Pizarski}.

\section{The character of the QCD phase transition.}

\begin{figure}[ht]	
\vskip -1.5 cm
\centerline{\epsfxsize 2.8 truein \epsfbox{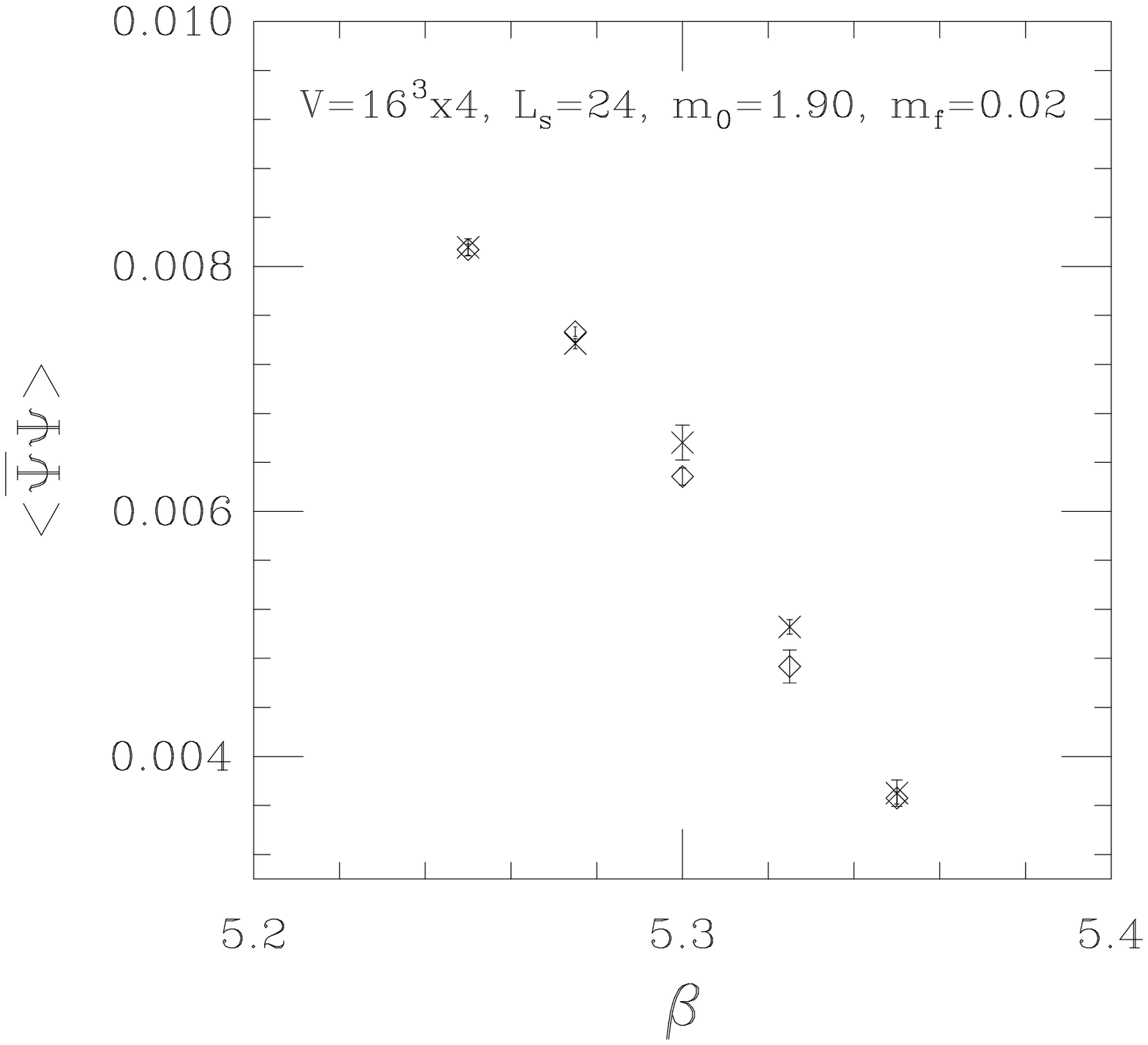}}   
\vskip -.2 cm
\caption[]{
\label{fig_6}
\small $<\Psibar \Psi>$ vs. $\beta$ on a $16^3 \times 4$ lattice with $m_0 = 1.9$, 
$m_f=0.02$ and $L_s=24$.}
\end{figure}

In order to determine the critical temperature and investigate the
width of the critical region and the order of the transition,
simulations of full QCD with DWF on $16^3 \times 4$ lattices close to
the transition are performed for $m_0=1.9$, $m_f=0.02$ and
$L_s=24$. The chiral condensate is plotted as a function of $\beta$ in
figure \ref{fig_6}.  For each $\beta$ two separate simulations are
done one with an ordered initial configuration (diamonds) and one with
a disordered one (crosses).

A zero temperature dynamical simulation on an $8^3 \times 32$ lattice
at $\beta = 5.325$ with $m_0=1.9$ $m_f=0.02$ and $L_s=24$ is done
in order to set the scale close to the transition.  The valence
extrapolated $\rho$ mass is found to be $m_\rho = 1.198(57)$.  If it is
used to set the scale it results to a critical temperature $T_c
\approx 161 {\rm MeV}$. The pion mass at $m_f=0.02$ is 
$m_\pi \approx 412 {\rm MeV}$. Further analysis attributes this large
value to finite $L_s$ effects. This is under current investigation.

Finally, from figure \ref{fig_6} one can see that the ordered and
disordered points at $\beta=5.325$ disagree even though each was
determined after $400$ thermalization sweeps were discarded.  These
small discrepancies visible for $\beta = 5.3$ and $5.325$
suggest the presence of a critical region and indicate the need for
collecting more statistics, a process currently underway. Also,
$m_\pi$ must be reduced before the order of the transition can be
identified. This is currently in progress.

\section{Conclusions.}

A novel lattice fermion regulator, domain wall fermions, was used to
study QCD thermodynamics. For the first time, this regulator offers
the possibility to control the notorious lattice chiral/flavor
asymmetries that have impeded numerical studies of QCD thermodynamics.
The $U(1)_A$ symmetry above the deconfining transition was found to be
broken only by a small amount. The critical temperature for the
transition was calculated and was found to be in agreement with
previous estimates. Detailed studies of the transition region are
currently underway with the hope that a consistent picture of the
order of the transition and the relatively small $U(1)_A$ breaking will
emerge. However, smaller pion masses are needed before any conclusions
can be drawn and such work is now underway.

\section{Acknowledgments.}

The simulations were done on the $400$ Gflops QCDSP supercomputer \cite{lat98_NHC}
at Columbia. This research was partially supported by DOE under grant 
\# DE-FG02-92ER40699 and for PV partially under grant \# NSF-PHY96-05199.


\end{document}